 \documentclass[preprint]{aastex6}

\renewcommand\thefootnote{\arabic{footnote}}

\fullcollaborationName{The Friends of AASTeX Collaboration}

\begin{document}


\title{Influence of Earth-Directed Coronal Mass Ejections on the Sun's Shadow Observed by the Tibet-III Air Shower Array}




\author{ 
M.~Amenomori\altaffilmark{1}, X.~J.~Bi\altaffilmark{2},
D.~Chen\altaffilmark{3}, T.~L.~Chen\altaffilmark{4},
W.~Y.~Chen\altaffilmark{2}, S.~W.~Cui\altaffilmark{5},
Danzengluobu\altaffilmark{4}, L.~K.~Ding\altaffilmark{2},
C.~F.~Feng\altaffilmark{6}, Zhaoyang~Feng\altaffilmark{2},
Z.~Y.~Feng\altaffilmark{7}, Q.~B.~Gou\altaffilmark{2},
Y.~Q.~Guo\altaffilmark{2}, H.~H.~He\altaffilmark{2},
Z.~T.~He\altaffilmark{5}, K.~Hibino\altaffilmark{8},
N.~Hotta\altaffilmark{9}, Haibing~Hu\altaffilmark{4},
H.~B.~Hu\altaffilmark{2}, J.~Huang\altaffilmark{2},
H.~Y.~Jia\altaffilmark{7}, L.~Jiang\altaffilmark{2},
F.~Kajino\altaffilmark{10}, K.~Kasahara\altaffilmark{11},
Y.~Katayose\altaffilmark{12}, C.~Kato\altaffilmark{13},
K.~Kawata\altaffilmark{14}, M.~Kozai\altaffilmark{13,15},
Labaciren\altaffilmark{4}, G.~M.~Le\altaffilmark{16},
A.~F.~Li\altaffilmark{17,6,2}, H.~J.~Li\altaffilmark{4},
W.~J.~Li\altaffilmark{2,7}, C.~Liu\altaffilmark{2},
J.~S.~Liu\altaffilmark{2}, M.~Y.~Liu\altaffilmark{4},
H.~Lu\altaffilmark{2}, X.~R.~Meng\altaffilmark{4},
T.~Miyazaki\altaffilmark{13},
K.~Munakata\altaffilmark{13}, T.~Nakajima\altaffilmark{13},
Y.~Nakamura\altaffilmark{13}, H.~Nanjo\altaffilmark{1},
M.~Nishizawa\altaffilmark{18}, T.~Niwa\altaffilmark{13},
M.~Ohnishi\altaffilmark{14}, I.~Ohta\altaffilmark{19},
S.~Ozawa\altaffilmark{11}, X.~L.~Qian\altaffilmark{6,2},
X.~B.~Qu\altaffilmark{20}, T.~Saito\altaffilmark{21},
T.~Y.~Saito\altaffilmark{22}, M.~Sakata\altaffilmark{10},
T.~K.~Sako\altaffilmark{23,14}, J.~Shao\altaffilmark{2,6},
M.~Shibata\altaffilmark{12}, A.~Shiomi\altaffilmark{24},
T.~Shirai\altaffilmark{8}, H.~Sugimoto\altaffilmark{25},
M.~Takita\altaffilmark{14}, Y.~H.~Tan\altaffilmark{2},
N.~Tateyama\altaffilmark{8}, S.~Torii\altaffilmark{11},
H.~Tsuchiya\altaffilmark{26}, S.~Udo\altaffilmark{8},
H.~Wang\altaffilmark{2}, H.~R.~Wu\altaffilmark{2},
L.~Xue\altaffilmark{6}, Y.~Yamamoto\altaffilmark{10},
K.~Yamauchi\altaffilmark{12}, Z.~Yang\altaffilmark{2},
A.~F.~Yuan\altaffilmark{4},
L.~M.~Zhai\altaffilmark{3}, H.~M.~Zhang\altaffilmark{2},
J.~L.~Zhang\altaffilmark{2}, X.~Y.~Zhang\altaffilmark{6},
Y.~Zhang\altaffilmark{2}, Yi~Zhang\altaffilmark{2},
Ying~Zhang\altaffilmark{2}, Zhaxisangzhu\altaffilmark{4}, and
X.~X.~Zhou\altaffilmark{7} \\ (The Tibet AS$\gamma$ Collaboration) }

\altaffiltext{1}{Department of Physics, Hirosaki University, Hirosaki 036-8561, Japan}
\altaffiltext{2}{Key Laboratory of Particle Astrophysics, Institute of High Energy Physics, Chinese Academy of Sciences, Beijing 100049, China}
\altaffiltext{3}{National Astronomical Observatories, Chinese Academy of Sciences, Beijing 100012, China}
\altaffiltext{4}{Department of Mathematics and Physics, Tibet University, Lhasa 850000, China}
\altaffiltext{5}{Department of Physics, Hebei Normal University, Shijiazhuang 050016, China}
\altaffiltext{6}{Department of Physics, Shandong University, Jinan 250100, China}
\altaffiltext{7}{Institute of Modern Physics, SouthWest Jiaotong University, Chengdu 610031, China}
\altaffiltext{8}{Faculty of Engineering, Kanagawa University, Yokohama 221-8686, Japan}
\altaffiltext{9}{Faculty of Education, Utsunomiya University, Utsunomiya 321-8505, Japan}
\altaffiltext{10}{Department of Physics, Konan University, Kobe 658-8501, Japan}
\altaffiltext{11}{Research Institute for Science and Engineering, Waseda University, Tokyo 169-8555, Japan}
\altaffiltext{12}{Faculty of Engineering, Yokohama National University, Yokohama 240-8501, Japan}
\altaffiltext{13}{Department of Physics, Shinshu University, Matsumoto 390-8621, Japan}
\altaffiltext{14}{Institute for Cosmic Ray Research, The University of Tokyo, Kashiwa 277-8582, Japan}
\altaffiltext{15}{Institute of Space and Astronautical Science, Japan Aerospace Exploration Agency (ISAS/JAXA), Sagamihara 252-5210, Japan}
\altaffiltext{16}{National Center for Space Weather, China Meteorological Administration, Beijing 100081, China}
\altaffiltext{17}{School of Information Science and Engineering, Shandong Agriculture University, Taian 271018, China}
\altaffiltext{18}{National Institute of Informatics, Tokyo 101-8430, Japan}
\altaffiltext{19}{Sakushin Gakuin University, Utsunomiya 321-3295, Japan}
\altaffiltext{20}{College of Science, China University of Petroleum, Qingdao 266555, China}
\altaffiltext{21}{Tokyo Metropolitan College of Industrial Technology, Tokyo 116-8523, Japan}
\altaffiltext{22}{Max-Planck-Institut f\"ur Physik, Munich D-80805, Germany}
\altaffiltext{23}{Escuela de Ciencias F\'{\i}sicas y Nanotechnolog\'{\i}a, Yachay Tech, Imbabura 100115, Ecuador}
\altaffiltext{24}{College of Industrial Technology, Nihon University, Narashino 275-8576, Japan}
\altaffiltext{25}{Shonan Institute of Technology, Fujisawa 251-8511, Japan}
\altaffiltext{26}{Japan Atomic Energy Agency, Tokai-mura 319-1195, Japan}


\begin{abstract}
We examine the possible influence of Earth-directed coronal mass ejections (ECMEs) on the Sun's shadow in the 3~TeV cosmic-ray intensity observed by the Tibet-III air shower (AS) array. We confirm a clear solar-cycle variation of the intensity deficit in the Sun's shadow during ten years between 2000 and 2009. This solar-cycle variation is overall reproduced by our Monte Carlo (MC) simulations of the Sun's shadow based on the potential field model of the solar magnetic field averaged over each solar rotation period. We find, however, that the magnitude of the observed intensity deficit in the Sun's shadow is significantly less than that predicted by MC simulations, particularly during the period around solar maximum when a significant number of ECMEs is recorded. The $\chi^2$ tests of the agreement between the observations and the MC simulations show that the difference is larger during the periods when the ECMEs occur, and the difference is reduced if the periods of ECMEs are excluded from the analysis. This suggests the first experimental evidence of the ECMEs affecting the Sun's shadow observed in the 3~TeV cosmic-ray intensity.
\end{abstract}




\section{Introduction} \label{sec:intro}

Coronal mass ejections (CMEs) are large magnetized clouds/blobs of plasma ejected by solar eruptions spreading into the interplanetary space. A fast CME forms/drives a strong shock and a magnetic sheath containing the strong and turbulent magnetic field behind the shock. Some CMEs also have rope-like magnetic structures behind the magnetic sheath \citep{Burlaga81, Klein82}. A CME propagating away from the Sun affects (the otherwise steady) galactic cosmic rays (GCRs) in a number of ways. The best-known consequence is the so-called ``Forbush decrease,'' which is observed when the detector enters a region of suppressed GCR density behind the shock driven by a CME \citep{Cane00}. It has been known that the CMEs sometimes may trigger major geomagnetic storms when they reach the Earth \citep{Gosling90, Gosling93}.

While {\it in-situ} measurements on board a satellite give the 1D distributions of the magnetic field in the CME along the satellite's path, it is still difficult to precisely derive the 3D large-scale magnetic structure of a CME. TeV and sub-TeV GCRs with large Larmor radii can sense the remote CMEs and give useful additional information on their structure. Observations of these high-energy GCRs are also relevant to space weather, because these particles travel nearly at the speed of light, sense the CME, and escape into the upstream region of the shock. They can easily overtake the much slower CME, bringing an advance warning of the Earth-directed CME (ECME) approaching the Earth \citep{Munakata00, Rockenbach14}. The solar modulation of sub-TeV GCRs has been already reported \citep{Munakata10}, but the influence of ECMEs has not yet been addressed above 1~TeV.

The Sun blocks high-energy GCRs arriving from the direction behind the Sun and casts a shadow (Sun's shadow) in the GCR intensity, which is influenced by the magnetic field of the Sun \citep{Clark57}. The small air shower (AS) array (Tibet-I) consisting of 45 scintillation detectors with a 15~m spacing started operation at Yangbajing (4300~m above sea level) in Tibet, China in 1990 and successfully observed the first evidence of the magnetic field's influence on the Sun's shadow in the 10 TeV energy range \citep{Amenomori93, Amenomori96}. Afterwards, the large AS array (Tibet-II) consisting of 221 scintillation detectors with the same 15 m spacing started operation in 1996 and successfully observed a clear solar-cycle variation of the Sun's shadow during a full activity-cycle from 1996 to 2009 \citep{Amenomori13}. To clarify the physical implications of the observed solar-cycle variation, detailed numerical simulations of the Sun's shadow have been developed based on the potential field models of the coronal magnetic field \citep{Wiegelmann15}. It was found that the simulated intensity deficit in the Sun's shadow is very sensitive to the coronal magnetic field models, and the observed solar-cycle variation is best reproduced by the current sheet source surface (CSSS) model. This was the first successful attempt of evaluating the coronal magnetic field models by using the Sun's shadow observed in the TeV GCR flux \citep{Amenomori13}. A higher density array with a 7.5 m spacing then started operation in late of 1999 \citep{Amenomori03}. This array (Tibet-III) can observe lower energy GCRs by lowering the energy threshold below 3~TeV \citep{Amenomori18}. In this paper, we analyze the Sun's shadow observed by Tibet-III, and report the influence of ECMEs on the Sun's shadow observed in the 3~TeV GCR intensity. \\

\section{Data Analysis} \label{sec:exp}

We analyze AS events observed from the direction of the Sun by Tibet-III in 2000-2009. The overall angular resolution and the modal energy of primary GCRs recorded with the array are estimated to be $0\fdg9$ and 3~TeV, respectively. We define the angular resolution as an angular radius containing 50\% of AS events produced by primary GCRs arriving from an identical direction.  We calculate the number of on-source events ($N_{\rm  on}$) as the number of events arriving from a given direction within a circular window centered at a certain point on the celestial sphere. The number of background or off-source events ($\langle N_{\rm off} \rangle$), on the other hand, is calculated as the average number of events within each of the eight off-source windows which are located at the same zenith angle as the on-source window, but apart by $\pm 9\fdg6$, $\pm 12\fdg8$, $\pm 16\fdg0$ and $\pm 19\fdg2$ in the azimuthal direction \citep{Amenomori09}.  The window radius of $0\fdg9$ (angular resolution for 3~TeV GCRs) is adopted for calculating both $N_{\rm on}$ and $\langle N_{\rm off} \rangle$. We calculate $N_{\rm on}$ and $\langle N_{\rm off} \rangle $ on each gridpoint on $0\fdg1 \times 0\fdg1$ mesh of the geocentric solar ecliptic (GSE) longitude and latitude surrounding the optical center of the Sun.  We then estimate the deficit relative to the number of background events as $D_{\rm obs}=( N_{\rm on} - \langle N_{\rm off} \rangle) / \langle N_{\rm off} \rangle$ at every grid. We also apply the same method to calculate $D_{\rm obs}$ in the Moon's shadow in the equatorial coordinate system.

\begin{figure}[ht!]
\epsscale{1.0}
\plotone{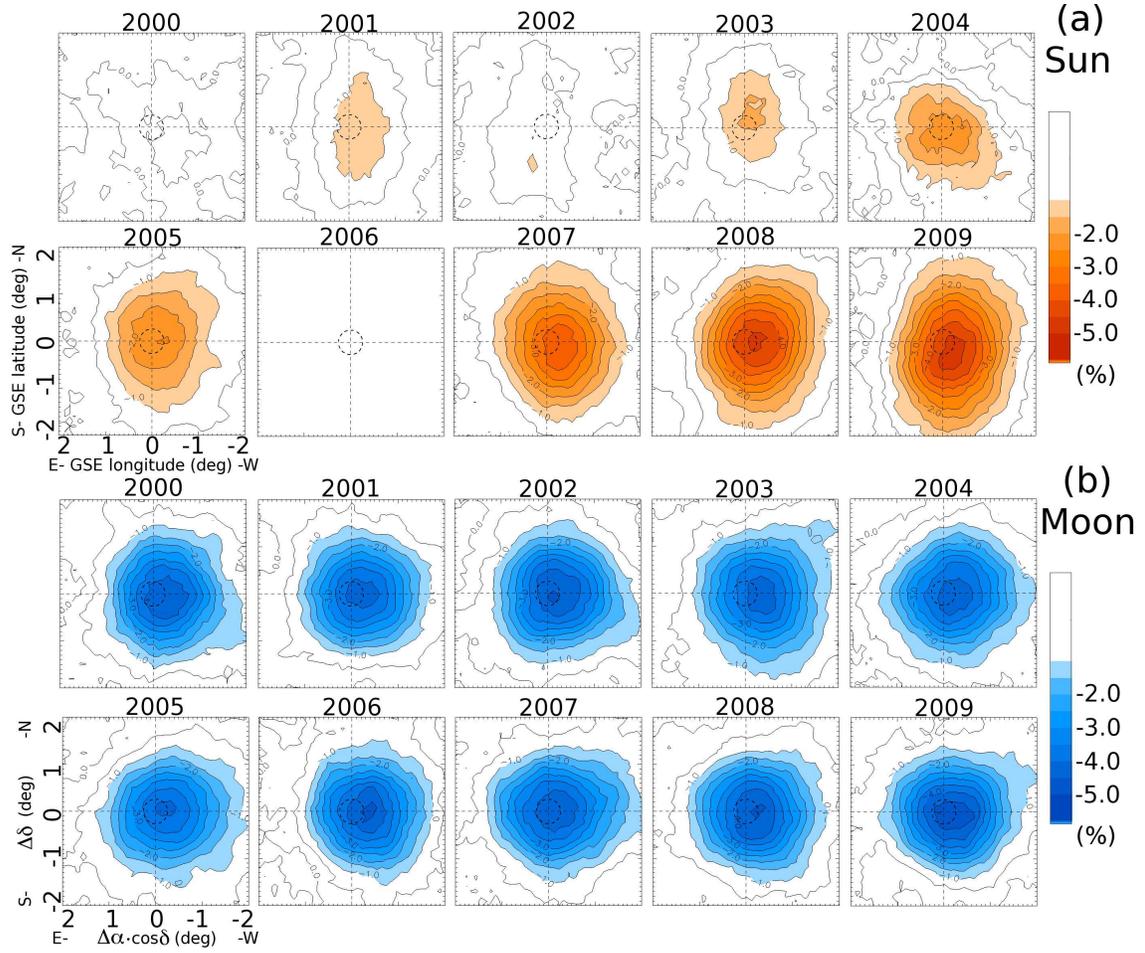}
\caption{Year-to-year variation of (a) the Sun's shadow and (b) Moon's shadow observed by the Tibet-III array between 2000 and 2009. The upper panels show 2D contour maps of $D_{\rm obs}$ in the Sun's shadow in the GSE coordinate system, while the lower panels display $D_{\rm obs}$ in the Moon's shadow each as a function of right ascension and declination relative to the apparent center of the Moon.
\label{fig1}}
\end{figure}

Figures~\ref{fig1} (a) and (b) show yearly maps of $D_{\rm obs}$ in the Sun's and the Moon's shadows, respectively, observed at 3~TeV in 2000-2009, except for 2006, which is excluded from Figure~\ref{fig1} (a) because of the insufficient statistics. The Moon's shadow provides us with a good reference of the detector stabilities, because the Moon has no magnetic field and its optical size viewed from the Earth is almost the same as the Sun \citep{Amenomori09}. It is seen in Figure 1 (a) that the Sun's shadow is darker (with larger negative $D_{\rm obs}$) around 2008 when the solar activity was close to the minimum, while it becomes faint (with smaller negative $D_{\rm obs}$) around 2001 when the activity was close to the maximum. In contrast, the observed Moon's shadow shown in Figure~\ref{fig1} (b) is quite stable, ensuring the instrumental stability during the same period.\\

\section{MC Simulation} \label{sec:mc}

For the primary GCRs in our Monte Carlo (MC) simulations of the Sun's shadow, we assume the energy spectra and the elemental compositions modeled by \citet{Shibata10} which compile various measurements between 0.3~TeV and 1000~TeV. We set the minimum energy of primary particles at 0.3~TeV, well below the threshold energy for triggering our AS detection. We perform the simulations as follows: (1) we first generate AS events randomly at the top of the atmosphere along the Sun's orbit using the CORSIKA code \citep{Heck98} with the EPOS-LHC interaction model \citep{Pierog15} so that AS cores uniformly distribute within 300~m radius from the center of array. This area within 300~m from the center sufficiently covers detectors actually hit by AS particles; (2) we then distribute these simulated events among detectors configuring Tibet-III by the GEANT4 code \citep{Agostinelli03} and calculate the output of each detector, which can be analyzed for AS reconstruction and event selections in the same way as the experimental data; (3) we assign an opposite charge to each primary particle at the top of the atmosphere and shoot it in random directions within $4^\circ$ from the Sun, each called as the initial shooting direction; (4) we trace the orbital motion of each particle back to the Sun in the model magnetic fields (see the next section) by the fourth-order Runge-Kutta method; (5) if a particle hits the photosphere, its initial shooting direction is regarded as the direction contributing to the intensity deficit in the Sun's shadow. After smearing initial shooting directions mimic the angular resolution event by event, we finally get the expected Sun's shadow equivalent to the observation \citep{Amenomori13}.\\

\section{Magnetic Field Model} \label{sec:mc}

For the solar magnetic field model in the MC simulation, we adopt the CSSS model which is the potential field model most successfully reproducing the temporal variation of the Sun's shadow observed with Tibet-II at 10 TeV \citep{Amenomori13}. The potential field models describe the coronal magnetic field based on the optical measurements of the  photospheric magnetic field. In our MC simulations, we use the photospheric field observed with the spectromagnetograph of the National Solar Observatory at Kitt Peak \citep{Jones92} in each Carrington rotation (CR) period ($\sim$27.3 days). The CSSS model \citep{Zhao95} involves four free parameters, the radius $R_{\rm ss}$ of the spherical source surface (SS) where the supersonic solar wind starts blowing radially, the order $n$ of the spherical harmonic series describing the observed photospheric field, the radius $R_{\rm cp}$ ($=1.7R_{\bigodot}$) of  the sphere where the magnetic cusp structure in the helmet streamers appears, and the length scale $l_{a}$ of horizontal coronal electric currents. In the present paper, we set $l_{a}$ to be one solar radius ($l_{a}=R_{\bigodot}$) and examine two different cases with $R_{\rm ss}$ =2.5$R_{\bigodot}$ and $R_{\rm ss}$ = 10$R_{\bigodot}$. The former $R_{\rm ss}$ is a standard value used in the original paper \citep{Zhao95}, while the latter gained support from some recent evidences \citep{Balogh95,Zhao02,Schussler06}. We set $n=10$ which is sufficient to describe fine structures relevant to the orbital motion of high-energy particles with large Larmor radii. The radial component of the coronal magnetic field at $R_{\rm ss}$ is then stretched out forming the Parker's spiral interplanetary magnetic field \citep[IMF;][]{Parker58}.  For the radial solar wind speed needed in the Parker's model, we use the solar wind speed synoptic chart estimated from the interplanetary scintillation measurement in each CR and averaged over the Carrington longitude \citep{Tokumaru10}.\footnote{\url{http://stsw1.isee.nagoya-u.ac.jp/ips_data-e.html}} We adopt a dipole model for the geomagnetic field.\\

\section{Results and Discussions} \label{sec:results}

\begin{figure}[ht!]
\epsscale{1.0}
\plotone{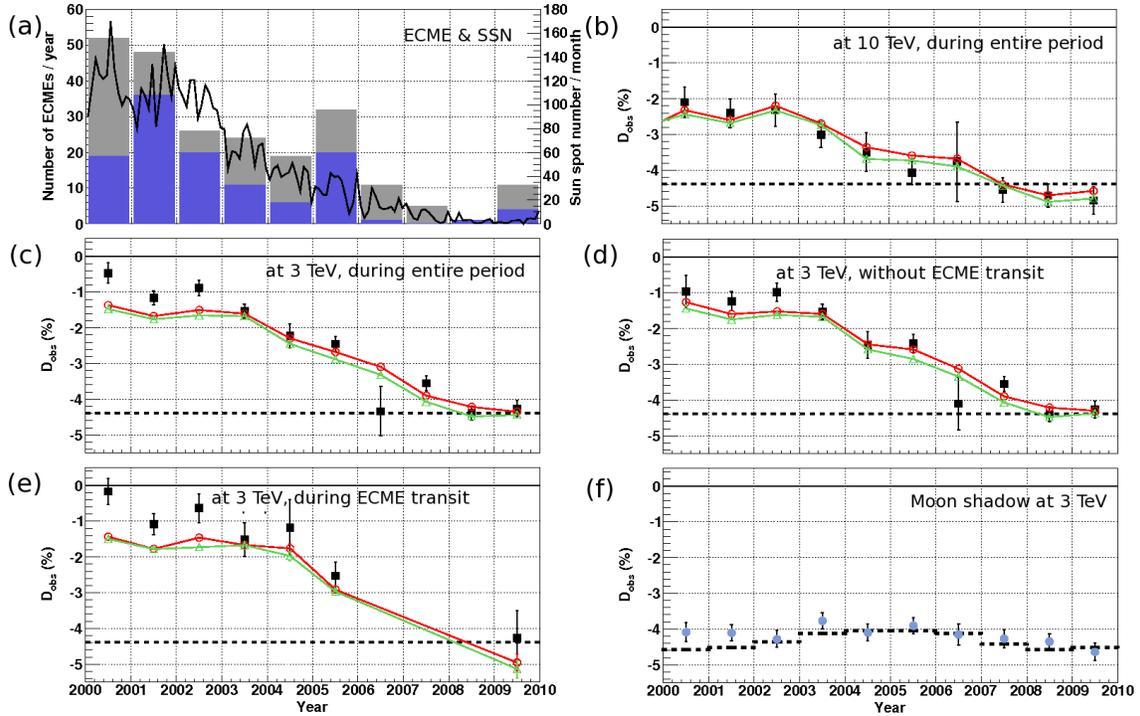}
\caption{
Temporal variations of $D_{\rm obs}$ in the Sun's shadow during ten years between 2000 and 2009. For reference, panel (a) shows the variations of the monthly mean sunspot number and the number of ECMEs recorded in each year plotted on the right and left vertical axes, respectively. The gray histogram in panel (a) shows all ECMEs listed in the catalog by \citet{Richardson10}, while the blue histogram displays only ECMEs each with the transit periods covered by the observation of the Sun's shadow by the Tibet-III array. The solid squares in each panel from (b) to (f) display (b) $D_{\rm obs}$ in the Sun's shadow observed at 10~TeV by the same detector configuration as the Tibet-II array \citep{Amenomori13}, (c) $D_{\rm obs}$ in the Sun's shadow observed by the Tibet-III array at 3~TeV during an entire period, (d) $D_{\rm obs}$ of the Sun's shadow at 3~TeV observed during a period without ECMEs transit periods, (e) $D_{\rm obs}$ of the Sun's shadow at 3~TeV observed during the ECME transit periods that are excluded in (d), and (f) $D_{\rm obs}$ of the Moon's shadow at 3~TeV. The error bars indicate the statistical errors. The dashed lines in panels (b)-(e) and a dashed curve in (f) indicate $D_{\rm obs}$ expected from the apparent angular size of the Sun and the Moon. The red open circles and green open triangles in panels (b)-(e) display $D_{\rm obs}$ predicted by two different MC simulations using the CSSS models with $R_{\rm ss}$ = 2.5$R_{\bigodot}$ and $R_{\rm ss}$ = 10.0$R_{\bigodot}$, respectively.
\label{fig2}}
\end{figure}

In the present paper, we analyze the temporal variation of $D_{\rm obs}$, measured at the center of the yearly mean 2D map in Figure~\ref{fig1} (a). For reference, the solid curve in Figure~\ref{fig2} (a) displays the monthly mean sunspot number representing the solar activity on the right vertical axis, while the solid squares in Figure~\ref{fig2} (b) show the temporal variation of $D_{\rm obs}$ observed by Tibet-II at 10 TeV. It is seen in Figure~\ref{fig2} (b) that the magnitude $|D_{\rm obs}|$ shows a clear solar-cycle variation decreasing with increasing solar activity. It is indicated by our MC simulations that, during the solar maximum, GCRs passing near the solar limb are scattered by the strong and complicated coronal magnetic field and may appear from the direction of the optical solar disc reducing $|D_{\rm obs}|$. Also shown in Figure~\ref{fig2} (b) by the red open circles and green open triangles are predictions by two different MC simulations using the CSSS models with $R_{\rm ss}$ = 2.5$R_{\bigodot}$ and $R_{\rm ss}$ = 10.0$R_{\bigodot}$, respectively. It is seen that these models successfully reproduce the temporal variation over an entire period \citep{Amenomori13}.

Solid squares in Figure~\ref{fig2} (c) show the temporal variation of $D_{\rm obs}$ observed at 3 TeV by Tibet-III. The solar-cycle variation is again clearly seen, but with a larger amplitude when compared with Figure~\ref{fig2} (b) at 10 TeV, due to the larger magnetic deflection expected for lower energy GCRs in the solar corona. This energy dependent feature of the solar-cycle variation is overall reproduced by the MC simulations using the CSSS models. We also note, however, that $|D_{\rm obs}|$ in Figure~\ref{fig2} (c) is significantly less than those predicted by MC simulations particularly in 2000-2002 around the solar maximum with the statistical significance 4.6$\sigma$ and 5.4$\sigma$ for the CSSS models with $R_{\rm ss}$ = 2.5$R_{\bigodot}$ and $R_{\rm ss}$ = 10.0$R_{\bigodot}$, respectively.  Such deviations from MC simulations are not seen in Figure~\ref{fig2} (b) for 10 TeV GCRs.

The $\chi^2$ test of the agreement between the observed and simulated $D_{\rm obs}$s in Figure~\ref{fig2} are summarized in Table~\ref{tab1}. The solid squares in Figure~\ref{fig2} (f) show the temporal variation of $D_{\rm obs}$ in the Moon's shadow observed by Tibet-III array, while the dashed curve indicates the variation of $D_{\rm obs}$ expected from the small variation of the distance between the Earth and the Moon. We estimate the systematic error of $D_{\rm obs}$ in Table~\ref{tab1} to be 0.178\% from the deviation of $D_{\rm obs}$ from the dashed curve in this figure. It is clear in Table~\ref{tab1} that the deviation of $D_{\rm obs}$ from the MC simulation during an entire period in Figure~\ref{fig2} (c) is exceeding the acceptable error range.

In order to examine the influence of the ECMEs that occur most frequently during the solar maximum period, we use the catalog of CMEs compiled by \citet{Richardson10} that includes 228 ECMEs recorded in 2000-2009, each with a given arrival time at the Earth. While an ECME arrives at the Earth $2-4$ days after the solar eruption, 3~TeV GCRs take only $\sim$8 minutes to reach the Earth after passing near the Sun. GCRs arriving at the Earth from the direction of the Sun, therefore, can be affected by the ECME during a transit period between the solar eruption time observed by the {\it Solar and Heliospheric Observatory} ({\it SOHO})/Large Aperture Solar Coronagraph (LASCO) and the arrival time at the Earth. In this study, we examine the influence of ECMEs by analyzing $D_{\rm obs}$ observed during a period with/without these transit periods of ECMEs. For ECMEs lacking relevant eruption time available from the SOHO/LASCO observation, we assume 4 days for the transit period, which is an average of all ECMEs covered by the SOHO/LASCO observation. 

The histograms in Figure~\ref{fig2} (a) show the number of ECMEs recorded in each year. The gray histogram displays all 228 ECMEs listed in Richardson \& Cane's catalog, while the blue histogram shows only 118 ECMEs each with a transit period covered by the observation of $D_{\rm obs}$ by Tibet-III. It is seen that the number of ECMEs varies roughly in a positive correlation with the sunspot number. Due to this correlation, almost half of the analysis period is occupied by $20-35$ ECMEs per year in 2000-2002 around the solar maximum, while more than 90\% of the analysis period contains no ECME in 2007-2009 around the solar minimum. The solid squares in Figure~\ref{fig2} (d) display $D_{\rm obs}$ observed during periods without ECMEs transit periods, while the solid squares in Figure~\ref{fig2} (e) show $D_{\rm obs}$ observed during the ECME transit periods that are excluded in Figure~\ref{fig2} (d). It is seen in these figures that the deviation of $D_{\rm obs}$ from the MC simulations in 2000-2002 is significantly reduced in Figure~\ref{fig2} (d) than in Figure~\ref{fig2} (c), while it is increased in Figure~\ref{fig2} (e), as confirmed by the $\chi^2$ tests in Table~\ref{tab1}. Interestingly, the $D_{\rm obs}$ observed at 10~TeV by the Tibet-II array during the ECME transit period also seem to be marginally inconsistent with the MC simulation at the chance probability of 0.044 as seen in this table.

\begin{figure}[ht!]
\epsscale{0.6}
\plotone{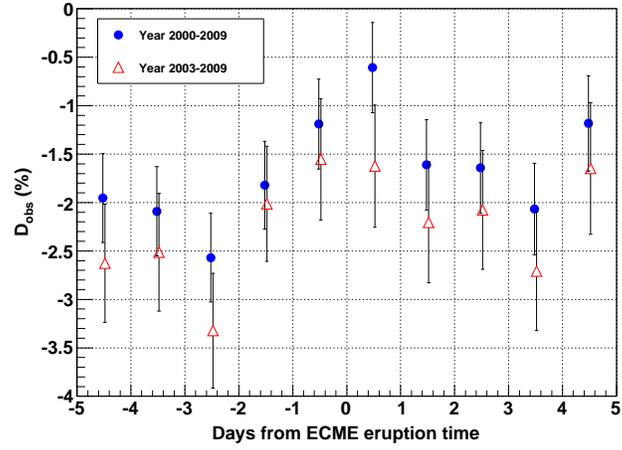}
\caption{
Superposed temporal variation of $D_{\rm obs}$ around the ECME eruption time set at $t=0$ on the horizontal axis. The blue solid circles (red triangles) show the variation obtained from the data in 2000-2009 (2003-2009).
\label{fig3}}
\end{figure}

Due to the insufficient statistics and the limited angular resolution, the current observation of the Sun's shadow with Tibet-III does not allow us to analyze each individual ECME event by event.  Hence, we superposed the temporal variations of $D_{\rm obs}$ on a daily basis with the CME eruption time set at $t=0$ in Figure~\ref{fig3}. In order to reduce the influence of the overlapping ECMEs, we excluded an ECME from this plot when it has another ECME eruption(s) within $\pm 5$ days from its eruption time. The blue solid circles show $D_{\rm obs}$ observed in the entire period including 2000-2002 when the transit periods were recorded with significant overlapping in the table by \citet{Richardson10}, while the red triangles show $D_{\rm obs}$ observed in 2003-2009 when the transit periods were less overlapping. Although the variations in this plot may not be statistically compelling, one can still see the tendency that $D_{\rm obs}$ has a maximum at $t=0$, and gradually decreases during the following couple of days. This may imply that the ECME continues to affect the Sun's shadow even after getting out to interplanetary space. This feature is better seen in the 2003-2009 period (red triangles) that is less contaminated by overlapping CMEs. We also note that $D_{\rm obs}$ starts to gradually increase before the ECME eruption ($t<0$). This again may suggest that the Sun's shadow can be a useful tool for space weather forecasting. It is also noted in this figure that the temporal variation of $D_{\rm obs}$ due to the ECME is separate from the solar-activity-cycle variation that appears in the difference between the background (average) levels in 2000-2009 and 2003-2009.

These results give a supporting evidence of the influence of the ECMEs on $D_{\rm obs}$ observed at 3~TeV. The magnetic deflection of GCR orbit in the CME is proportional to $ L/R_{\rm L}$, where $L$ is the scale length of the magnetic field and $R_{\rm L}$ is the Larmor radius of GCRs. Due to the self-similar expansion of the CME, $L$ increases with the radial distance $r$ from the Sun as $L \propto r$, while $R_{\rm L}$ also increases as $R_{\rm L} \propto r^{2}$ because of the transverse field component, $B_{\rm t}$ responsible to the orbital deflection, decreasing as $B_{\rm t} \propto 1/r^{2}$ in the expanding magnetic flux rope. So, the deflection power of the CME varies as $1/r$. The same deflection is, however, more effective in reducing $D_{\rm obs}$ when it occurs farther from the Sun, resulting in another factor proportional to $r$. The net influence on $D_{\rm obs}$, therefore, may remain nearly constant during the transit period of the ECME. If we assume $B_{t} = 10$~nT and $L = 0.1$~au at 1 au, we obtain $\sim2^{\circ}$ for the deflection angle which is larger than the optical diameter ($\sim 0\fdg5$) of the Sun viewed from the Earth and enough to decrease $D_{\rm obs}$.

As $R_{\rm L}$ of 3 TeV cosmic rays is as large as 0.1 $R_{\bigodot}$ even in a 1 G magnetic field, the Sun's shadow is expected to be more sensitive to the magnetic field with $L$ comparable to (or larger than) $R_{\rm L}$ and less sensitive to the smaller scale field near the photosphere, even if the field is stronger. $|D_{\rm obs}|$ is, therefore, expected to decrease significantly when such a large-scale magnetic field appears over the active regions on the solar limb. In order to check how a small-scale field in the solar corona can affect $D_{\rm obs}$, we performed the MC simulations with the CSSS model by increasing the maximum order of the harmonic series from $n=10-30$. The smaller scale in coronal magnetic field is expressed by the higher order harmonic component. As a result, we found that $D_{\rm obs}$ is insensitive to the smaller scale field with $n>10$. Since the $n$th order harmonic component of the field at $r$ is proportional to $(R_{\bigodot}/r)^{(n+2)}$, a smaller and stronger field on the photosphere represented with a larger $n$ diminishes faster with $r$ and cannot occupy a large volume ($\sim$$L^3$) sufficient to deflect TeV cosmic-ray orbits. We also performed our simulations by changing the photospheric magnetic field observations at Kitt Peak (Synoptic Optical Long-term Investigations of the Sun (SOLIS)/Kitt Peak Vacuum Telescope (KPVT)) to those by the Michelson Doppler Imager (MDI),\footnote{\url{http://soi.stanford.edu}} and Global Oscillation Network Group (GONG),\footnote{\url{https://gong.nso.edu}} which are known to have the different spatial resolution and dynamic range of the observed field strength \citep{Riley14}.  The average strength of the photospheric magnetic field by the MDI (GONG) observation is $1.5-2$ times higher (lower) than that of SOLIS/KPVT, but we found no significant difference in the calculated solar-cycle variations of $D_{\rm obs}$. This indicates that the strong magnetic fields around the active regions on the photosphere, which largely contribute to the average field strength, are mostly closed in the lower corona and incapable of occupying a large volume sufficient to affect $D_{\rm obs}$. These results indicate that $D_{\rm obs}$ is insensitive to small-scale magnetic structures in the potential field model.

The discussion above is only qualitative and we need the detailed magnetic field models of the ECME for quantitative studies. It is noted in Figures~\ref{fig2} (c)-(e) that the prediction of our MC simulation is almost insensitive to the ECME transit periods. This is because the potential field models used in our simulation are capable of reproducing only the average magnetic field in each CR period and incapable of properly representing the dynamic variations of the field in association with the ECMEs. \citet{Shiota10} and \citet{Den15} have recently developed the magnetohydrodynamic (MHD) simulations to reproduce the propagation and evolution of the ECMEs in the Sun$-$Earth space aiming at the space weather prediction. In the future, we will incorporate the magnetic fields by these MHD simulations into our MC simulations of the Sun's shadow. Comparisons of such simulations with the observed Sun's shadow will provide us with an alternative tool for evaluating the MHD models and for improving the space weather prediction. Due to the insufficient statistics and the limited angular resolution, the current observation of the Sun's shadow with Tibet-III does not allow us to analyze each individual ECME event by event. The future huge AS arrays will enable us to analyze the Sun's shadow during an individual ECME transit period on a daily or even on hourly basis.\\

\section{Summary} \label{sec:summary}
We confirmed a clear solar-cycle variation of the Sun's shadow observed at 3~TeV by Tibet-III in 2000-2009. By comparing this observed variation with the MC simulations, assuming the CSSS models for the coronal magnetic field, we find that the magnitude ($|D_{\rm obs}|$) of GCR intensity deficit in the Sun's shadow is significantly less than those of the MC simulations, particularly during years around the solar maximum when a significant number of ECMEs are recorded at the Earth. The $\chi^2$ test of the agreement between the observations and the MC simulations shows that $|D_{\rm obs}|$ is considerably increased by excluding the ECME transit periods from the analysis, while it is reduced in the analyses of the ECME transit periods. This is the first experimental result indicating the influence of the ECMEs on the Sun's shadow observed at 3~TeV.\\

\acknowledgments
The collaborative experiment of the Tibet AS arrays has been performed under the auspices of the Ministry of Science and Technology of China (No. 2016YFE0125500) and the Ministry of Foreign Affairs of Japan. This work was supported in part by a Grant-in-Aid for Scientific Research on Priority Areas from the Ministry of Education, Culture, Sports, Science, and Technology, by Grants-in-Aid for Science Research from the Japan Society for the Promotion of Science in Japan, and by the National Natural Science Foundation of China (Nos. 11533007 and 11673041) and the Chinese Academy of Sciences and the Key Laboratory of Particle Astrophysics, Institute of High Energy Physics, CAS. This work is also supported by the joint research program of the Institute for Cosmic Ray Research (ICRR), the University of Tokyo. K.~K. is supported by the Toray Science Foundation.  The authors thank Dr. Xuepu Zhao of Stanford University for providing the usage of the CSSS model. The authors thank Dr. J\'{o}zsef K\'{o}ta of University of Arizona for his useful comments and discussions.\\

\newpage

\begin{deluxetable}{rcccccccc}
\tabletypesize{\scriptsize}
\tablewidth{0pt} 
\tablenum{1}
\tablecaption{Summary of the $\chi^2$ tests of the agreement between $D_{\rm obs}$ and predictions by MC simulations based on the statistical and systematic errors \label{tab1}}
\tablehead{
\colhead{} & \multicolumn{2}{c}{During an Entire Period} & \colhead{} & \multicolumn{2}{c}{Without ECME Transit Periods} & \colhead{} & \multicolumn{2}{c}{During ECME Transit Periods}  \\
\cline{2-3}\cline{5-6}\cline{8-9}
\colhead{Models} &
\colhead{$\chi^2$/d.o.f.} & \colhead{Probability} & \colhead{} &
\colhead{$\chi^2$/d.o.f.} & \colhead{Probability} & \colhead{} &
\colhead{$\chi^2$/d.o.f.} & \colhead{Probability} 
}
\startdata
3~TeV CSSS $R_{\rm ss}$=2.5$R_{\bigodot}$ & 21.3(32.1)/10 & 0.019($3.9\times10^{-4}$) &&  8.6(12.2)/10 & 0.57(0.27) &&  19.2(23.9)/7 & 0.0076(0.0012) \\
3~TeV CSSS $R_{\rm ss}$=10$R_{\bigodot}$ & 30.1(46.9)/10 & $8.2\times10^{-4}$($9.8\times10^{-7}$) && 14.3(21.0)/10 & 0.16(0.021) &&  24.0(29.4)/7 & 0.0011($1.2\times10^{-4}$) \\
\hline
$^{a}$10~TeV CSSS $R_{\rm ss}$=10$R_{\bigodot}$ & 8.3(10.3)/14 & 0.87(0.74) &&  7.5(8.9)/14 & 0.91(0.84) && 20.1(21.1)/11 & 0.044(0.032) \\
\enddata
{\bf Note.} Results based on only the statistical errors are indicated in braces.
\tablenotetext{a}{For $D_{\rm obs}$ observed at 10 TeV by the same detector configuration as the Tibet-II array during a period between 1996 and 2009 \citep{Amenomori13}.}
\end{deluxetable}

\newpage

\newpage

\newpage

\newpage

\end{document}